\documentclass[nohyperref]{article}

\usepackage[T1]{fontenc}

\usepackage{microtype}
\usepackage{graphicx}
\usepackage[dvipsnames,table,xcdraw]{xcolor}
\usepackage{subfigure}
\usepackage{booktabs} 

\usepackage{hyperref}

\newcommand{\bth}{\boldsymbol{\theta}
}
\newcommand{\mcal}[1]{\mathcal{#1}}
\newcommand{\bz}{\mathbf{z}}
\newcommand{\by}{\mathbf{y}}
\newcommand{\bx}{\mathbf{x}}

\newcommand{\bw}{\mathbf{w}}
\newcommand{\bTh}{\boldsymbol{\Theta}
}

\usepackage[accepted]{icml2022}

\usepackage{amsmath}
\usepackage{amssymb}
\usepackage{mathtools}
\usepackage{amsthm}
\renewcommand{\mid}{\,|\,}

\usepackage[capitalize,noabbrev]{cleveref}

\usepackage[acronym,smallcaps,nowarn,section,nogroupskip,nonumberlist,shortcuts]{glossaries}
\glsdisablehyper
\newacronym{sbi}{sbi}{simulation-based inference}
\newacronym{npe}{npe}{neural posterior estimation}
\newglossaryentry{abm}
{
  name={\textsc{abm}},
  description={agent-based model},
  first={\glsentrydesc{abm} (\glsentrytext{abm})},
  plural={\textsc{abm}s},
  descriptionplural={agent-based models},
  firstplural={agent-based models (\glsentryplural{abm})}
} 
\newacronym{maf}{maf}{masked autoregressive flow}
\newacronym{abc}{abc}{approximate Bayesian computation}

\newglossaryentry{rnn}
{
  name={\textsc{rnn}},
  description={recurrent neural network},
  first={\glsentrydesc{rnn} (\glsentrytext{rnn})},
  plural={\textsc{rnn}s},
  descriptionplural={recurrent neural networks},
  firstplural={\glsentrydescplural{rnn} (\glsentryplural{rnn})}
} 
\newglossaryentry{gnn}
{
  name={\textsc{gnn}},
  description={graph neural network},
  first={\glsentrydesc{gnn} (\glsentrytext{gnn})},
  plural={\textsc{gnn}s},
  descriptionplural={graph neural networks},
  firstplural={\glsentrydescplural{gnn} (\glsentryplural{gnn})}
} 
\newacronym{gru}{gru}{gated recurrent unit}

\theoremstyle{plain}

\theoremstyle{definition}

\theoremstyle{remark}

\usepackage[textsize=tiny]{todonotes}

\icmltitlerunning{
Calibrating Agent-based Models to Microdata with Graph Neural Networks}

\begin{document}

\twocolumn[
\icmltitle{
Calibrating Agent-based Models to Microdata with Graph Neural Networks}



\icmlsetsymbol{equal}{*}

\begin{icmlauthorlist}
\icmlauthor{Joel Dyer}{inet,mi}
\icmlauthor{Patrick Cannon}{imp}
\icmlauthor{J. Doyne Farmer}{inet,mi,sfi}
\icmlauthor{Sebastian M. Schmon}{imp,dur}
\end{icmlauthorlist}

\icmlaffiliation{inet}{Institute for New Economic Thinking, Oxford}
\icmlaffiliation{mi}{Mathematical Institute, University of Oxford}
\icmlaffiliation{imp}{Improbable, London}
\icmlaffiliation{sfi}{Santa Fe Institute}
\icmlaffiliation{dur}{Durham University}

\icmlcorrespondingauthor{Joel Dyer}{joel.dyer@maths.ox.ac.uk}

\icmlkeywords{Machine Learning, ICML}

\vskip 0.3in
]



\printAffiliationsAndNotice{}  

\begin{abstract}
Calibrating \glspl{abm} to data is among the most fundamental requirements to ensure the model fulfils its desired purpose. 
In recent years, \gls{sbi} methods have emerged as powerful tools for performing this task when the model likelihood function is intractable, as is often the case for \glspl{abm}.
In some real-world use cases of \glspl{abm}, both the observed data and the \gls{abm} output consist of the agents' states and their interactions over time. 
In such cases, there is a tension between the desire to make full use of the rich information content of such granular data on the one hand, 
and the need to reduce the dimensionality of the data to prevent difficulties associated with high-dimensional learning tasks on the other. 
A possible resolution is to construct lower-dimensional time-series through the use of summary statistics describing the macrostate of the system at each time point. However, a poor choice of summary statistics can result in an unacceptable loss of information from the original dataset, dramatically reducing the quality of the resulting calibration.
In this work, we instead propose to learn parameter posteriors associated with granular microdata directly using temporal graph neural networks. We will demonstrate that such an approach offers highly compelling inductive biases for Bayesian inference using the raw \gls{abm} microstates as output.
\end{abstract}

\section{Introduction}

\glsresetall

\Glspl{abm} are becoming a popular modelling tool in a variety of disciplines, from economics \citep{Baptista2016} to epidemiology \citep{ferguson2020impact}. They offer domain experts a high degree of flexibility in modelling complex systems, for example by naturally incorporating interactions between, and heterogeneity across, agents in the system. 

Typically, \glspl{abm} are stochastic, dynamic models in which the states $\bz^t = (\bz^t_{1}, \dots, \bz^t_{N})$ of a set of $N$ interacting agents, labelled $i = 1, \dots, N$, are simulated over time $t \in \left[0,T\right]$. We assume here that the \gls{abm} progresses in discrete\footnote{We discuss in Section \ref{sec:discuss} how this assumption may be relaxed.} timesteps $t=0, 1, \dots, T$, and that the agent states may be multidimensional such that $\bz^t_{i} \in \mathbb{R}^K$ for some $K \geq 1$. 
\Glspl{abm} further rely on a potentially time-varying graph structure -- represented as an adjacency matrix $\bw^t \in \mathbb{R}^{N\times N}$ -- that reflects, for example, the strength of the relationship between pairs of agents, or the set of pairwise interactions that can take place during the simulation. 
%
Once a set of parameters $\bth \in \bTh \subset \mathbb{R}^D$ and the initial states $\bz^0$ and $\bw^0$ are specified, the agent behaviours and interactions are simulated, and a time-series $\bx = (\bx^1, \dots, \bx^T)$ is generated as output. 
Typically, $\bx^t \in \mathbb{R}^M$ for some $M \geq 1$.

The model output $\bx$ is often taken to be some aggregate statistics describing the macrostate of the \gls{abm} over time, that is $\bx=g(\bw, \bz)$ for some aggregation function $g$.
In some cases, this is done out of necessity: it is sometimes the case that only aggregate data is available from the real-world system the \gls{abm} is designed to mirror and, consequently, the \gls{abm} can only be compared to reality through the lens of this aggregate data. Under these circumstances, two natural inference tasks arise:
\begin{enumerate}
    \item parameter inference -- i.e.~inferring the fixed parameters $\bth$; and
    \item latent state inference (filtering and smoothing) -- i.e.~inferring either or both the latent states $\bz^t$ and agent-agent relationships $\bw^t$ of and between the agents in the system over time.
\end{enumerate}
Both tasks are complicated by the fact that the relevant marginal and conditional likelihood functions are in general unavailable to compute, due to the complexity of \glspl{abm}. 
Intractable likelihoods are encountered widely across model types and application domains and, consequently, significant research effort within the statistics and machine learning communities has been directed towards developing likelihood-free, \gls{sbi} procedures that act as more convenient substitutes to their likelihood-based counterparts. 
For parameter inference, approaches such as \gls{abc} \citep{pritchard1999population, beaumont2002approximate, dyer2021approximate} have seen significant success, while more modern neural network 
approaches to density \citep{Papamakariosepsilon, Lueckmann2017, Greenberg2019} and density ratio \citep{thomas2021lfire, pmlr-v119-hermans20a} 
estimation show promise as a means to dramatically reduce the simulation burden in \gls{sbi} procedures; see \citet{dyer2022black} for a recent overview of these methods and their application to \glspl{abm} in the social sciences. 
Similarly, variants of the Kalman filter 
and sequential Monte Carlo methods 
have been developed and applied to the problem of \gls{abm} latent state inference \citep[see e.g.][]{ward2016dynamic, LUX2018391}, although this has received less attention within the \gls{abm} community than the problem of parameter inference.

In contrast to this formulation, the increasing availability of granular, longitudinal microdata on agent behaviours and interactions in real-world social systems (e.g. on social media) raises the possibility of dispensing with the structure described above, in which it is often necessary to perform two inference tasks to properly ``fit'' the \gls{abm}.
Indeed, in some crucial applications, the \gls{abm} is ``fully-observed'', i.e. $g(\cdot)=identity(\cdot)$ and $\bx^t=(\bw^t, \bz^t)$. 
Under these circumstances, the filtering and smoothing problems vanish and only the problem of parameter inference remains. In such cases, the process of fully calibrating the \gls{abm} to observed data is greatly simplified as a result of the granularity of the data. 

Approaches to performing parameter inference for fully-observed \glspl{abm} are currently lacking in  
the \gls{sbi} literature. Importantly, modellers are lacking approaches to \gls{sbi} that incorporate useful inductive biases that reflect the natural dynamic graph structure of the \gls{abm} and the data. The absence of such methods prevents us from properly capitalising on the availability, and full information content, of such granular data. 

In this paper, we address this gap by demonstrating how (recurrent) \glspl{gnn} may be combined with neural \gls{sbi} procedures to flexibly and automatically accommodate high-dimensional, high-resolution data describing the evolution of the microstate of a social system. We show that \glspl{gnn} provide useful inductive biases for the use of such microdata as observables against which parameters $\bth$ are calibrated in the case of ``fully observed'' \glspl{abm}, with promising performance on test cases modelling the coevolution of opinions and network structure in a social system. 

\section{Background \& Motivation}

\subsection{Simulation-based Parameter Inference}

\glsreset{sbi}

\Gls{sbi} is a set of algorithms in which likelihood-based parameter inference procedures -- such as Bayesian inference -- are approximated through training on \emph{iid} data $(\bx, \bth) \sim p(\bx \mid \bth) \pi(\bth)$, where $\pi(\bth)$ is a prior density and $p(\bx \mid \bth)$ is the likelihood function associated with the simulation model. 
This is done by first sampling $\bth \sim \pi(\bth)$ and subsequently forward simulating from the simulator at $\bth$, represented as $\bx \sim p(\bx \mid \bth)$.
%
Once trained, \gls{sbi} algorithms then, given some 
observation $\by$, yield estimates of parameter posteriors $\pi(\bth \mid \by)$ \citep{Papamakariosepsilon, Lueckmann2017, Greenberg2019}, data likelihood functions $p(\by \mid \bth)$ \citep{papamakarios2019sequential}, or likelihood-to-evidence ratios $p(\by \mid \bth)/p(\by)$ \citep{thomas2021lfire, pmlr-v119-hermans20a, dyer2022amortised}. 

Of particular interest to the current work is the first of these three alternatives, commonly referred to as \gls{npe}. In \gls{npe} algorithms, a conditional density estimator -- such as a mixture density network \citep{bishop1994mixture} or a normalising flow \citep{tabak2010density, tabak2013family, rezende} -- is trained to approximate the map $\bx \mapsto \pi(\cdot \mid \bx)$ using \emph{iid} training data $(\bx, \bth) \sim p(\bx \mid \bth) \pi(\bth)$. This provides the experimenter with immediate access to the posterior density estimated by the neural network, which can furthermore be constructed to operate directly on raw data $\bx$ through the incorporation of appropriate inductive biases into the network architecture. 

\begin{figure*}[tb]
    \centering
    \includegraphics[width=0.99\linewidth]{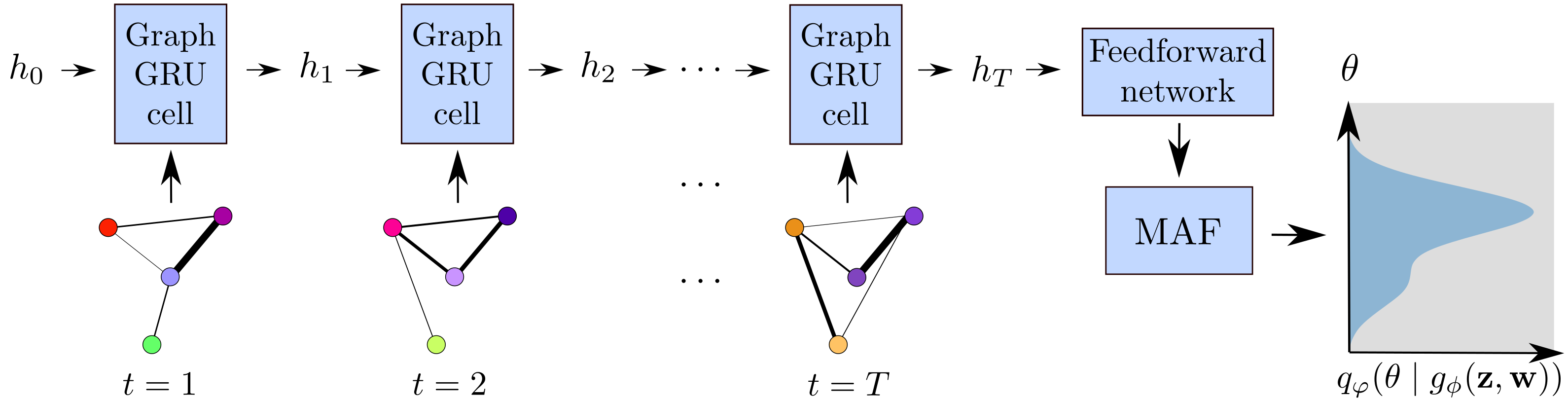}
    \caption{A schematic of the posterior estimation pipeline we use. The \gls{abm} -- shown as the dynamic graph with evolving node states (node colors) $\bz$ and edge weights (line widths) $\bw$ -- is embedded into a low-dimensional space with a graph \gls{gru} and a feedforward network applied to the \gls{gru}'s final hidden state, $h_T$. This representation, $g_{\phi}(\bz, \bw)$, is fed to the \gls{maf} to estimate the posterior as $q_{\varphi}(\bth \mid g_{\phi}(\bz, \bw))$.}
    \label{fig:setup}
\end{figure*}

\subsection{Graph Neural Networks}

\glsreset{gnn}

Recent years have seen considerable progress in the development of \glspl{gnn} in machine learning \citep[e.g.][]{yao2019graph, Zhang2020Efficient, BaekKH21}. In many cases, the design of a \gls{gnn} consists of generalising a convolution operator from regular, Euclidean domains -- as appears in convolutional neural networks 
-- to graphs. This has predominantly proceeded by constructing a convolution in the spatial domain \citep[see e.g.][]{Masci2015Geo, niepert2016learning} or by exploiting the convolution theorem and performing a multiplication in the graph Fourier domain \citep[see e.g.][]{bruna2014spectral}. A recent review of \glspl{gnn} and their design can be found in \citet{ZHOU202057}.

The problem of extending \glspl{gnn} to dynamic graphs has also recently received significant attention. In this vein, \citet{li2017diffusion} introduce Diffusion Convolutional Recurrent Neural Networks, with applications to traffic flow prediction. In addition, \citet{seo2018structured} propose Graph Convolutional Recurrent Networks, an adaptation of standard recurrent networks to operate on sequences of graphs via graph convolutional operators. Further examples of recurrent graph neural network architectures exist; a broader survey of neural networks for dynamic graphs can be found in \citet[][Section 7]{wu2021comp}.

\section{Methods}

In this paper, we consider the problem of parameter inference for the case of fully observed \glspl{abm}, where the data the experimenter observes is the complete trace of agent states $\bz^t$ and their relationships $\bw^t$ over all timesteps $t = 0, \dots, T$. Under these circumstances, 
the experimenter requires approaches to parameter inference that accommodate the 
dynamic graph structure of this data.

To address this, we construct a neural posterior estimator in which a recurrent \gls{gnn} $g_{\phi}$ and a neural conditional density estimator $q_{\varphi}$ are paired to approximate the map $(\bz, \bw) \mapsto \pi( \cdot \mid \bz, \bw)$, where $\phi$ and $\varphi$ are the parameters of the respective neural networks. In particular, we take $g_{\phi}$ to be a feedforward network applied to the final hidden state of the recurrent \gls{gnn} and $q_{\varphi}$ to be a normalising flow. The posterior may then be learned from this low-dimensional representation of the original high-dimensional sequence of graphs as $q_{\varphi}(\cdot \mid g_{\phi}(\bz, \bw))$.

While many choices of architecture are available, we employ a graph convolutional \gls{gru} \citep{seo2018structured} to construct $g_{\phi}$ and use a \gls{maf} \citep{papamakarios2017masked} for $q_{\varphi}$ in this paper. In Figure \ref{fig:setup}, we show a schematic of the pipeline that results from the particular choice of recurrent \gls{gnn} and conditional density estimator used for our experiments, although the exact modules appearing in this experimental setup may be substituted for others without fundamentally altering the overall pipeline. We train the network parameters $\phi$ and $\varphi$ concurrently and on the same loss function as the \gls{maf}, such that the graph sequence embedding and the posterior are learned simultaneously. Further details on the architecture and training procedure we employ can be found in the supplement. 

\section{Experimental Results}

To test the approach, we consider a task based on the Hopfield model of social dynamics proposed by \citet{macy2003polarization} which describes the coevolution of opinions and the social network structure, and the emergence of polarisation, in a population of $N$ agents. At each time step $t = 1, \dots, T$, each agent is equipped with $N-1$ undirected ties to the remaining agents in the population, and the strength and valence of the tie between agents $i$ and $j$ is characterised by $w^t_{ij} \in \left[-1, 1\right]$. Each agent is also equipped with a state vector $\bz^t_i \in \lbrace{-1, 1\rbrace}^K, i = 1, \dots, N$, which may represent the opinion status of agent $i$ on each of a number $K \geq 1$ of topics at time $t$. The \emph{social pressure} that agent $i$ experiences on topic $k$ at time $t$ is then modelled as
\begin{equation}
    P^t_{ik} = \frac{1}{N-1} \sum_{j \neq i} w^t_{ij} z^t_{ik},
\end{equation}
and $i$'s corresponding propensity to adopt the positive opinion is taken to be
\begin{equation}
    \pi^t_{ik} = \frac{1}{1 + e^{-\rho \cdot P^t_{ik}}},
\end{equation}
where $\rho > 0$ is a free parameter of the model. Agent $i$ then adopts the positive opinion on topic $k$ at time $t$ (i.e. $z^{t+1}_{ik} = 1$) if
\begin{equation}
    \pi^t_{ik} > 0.5 + \epsilon U^t_i,
\end{equation}
where $\epsilon \in \left[0, 1\right]$ is a further free parameter of the model and $U_i^t \sim \mathcal{U}(-0.5,0.5)$. Finally, the ties between agents evolve as
\begin{equation}
    w^{t+1}_{ij} = (1 - \lambda)w^t_{ij} + \frac{\lambda}{K} \sum_{k=1}^K z^{t+1}_{ik} z^{t+1}_{jk},
\end{equation}
where $\lambda \in \left[0,1\right]$ is a third free parameter of the model. 

Taking the initial proportion $p \in \left[0,1\right]$ of all opinion state entries as the final free parameter of the model, we assume the goal of approximating a posterior density for $\bth = (\rho, \epsilon, \lambda, p)$. Specifically, we assume prior densities $\rho \sim \mcal{U}(0,5)$, $\epsilon \sim \mcal{U}(0,1)$, $\lambda \sim \mcal{U}(0,1)$, $p \sim \mcal{U}(0,1)$, and further assume that the \gls{abm} is fully observed in the sense that all $w_{ij}^t$ and $z^t_{ik}$ are observed. To generate a pseudo-true dataset, we simulate the model for 25 time steps with $N=20$ at ground-truth parameter values $\bth^{*} = (1, 0.8, 0.5, 0.5)$.

\begin{figure}
    \centering
    \includegraphics[width=\linewidth]{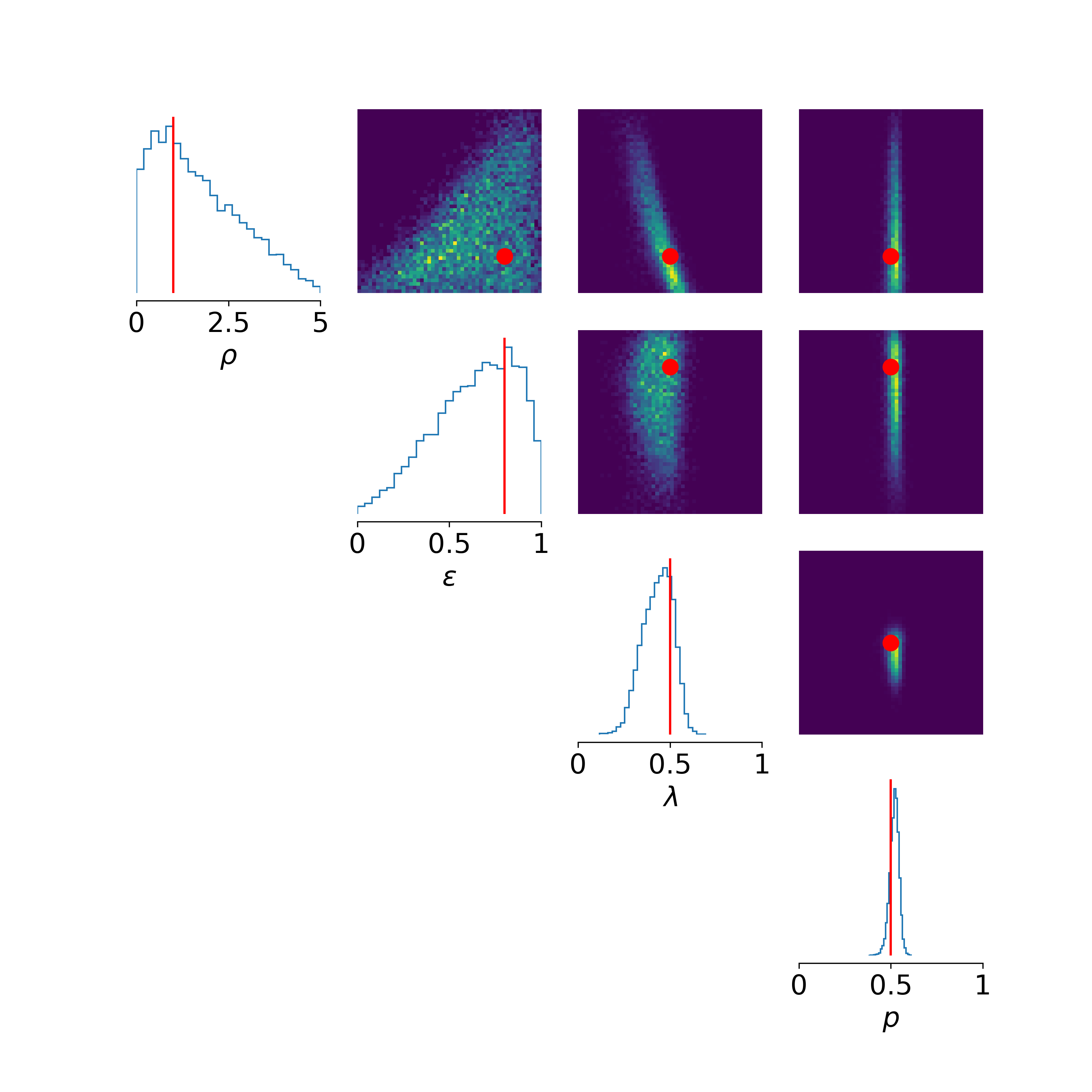}
    \vspace{-10mm}
    \caption{Approximate posterior for the Hopfield model obtained with a masked autoregressive flow and recurrent graph embedding network. Red lines/points show the ground truth parameters for the dataset.}
    \label{fig:Hopfield_MAF}
\end{figure}

We show the approximate posterior obtained with a \gls{maf} posterior estimator, recurrent graph convolutional embedding network, and a budget of 1000 simulations in Figure \ref{fig:Hopfield_MAF}. The diagonal subplots show the marginal posterior densities, while the off-diagonal subplots show the 2-dimensional projections of the joint densities. We show the ground truth parameters $\bth^{*}$ with red lines and points. The approximated posterior assigns high posterior density to the ground truth parameters, providing evidence that a reasonable degree of accuracy has been achieved by the posterior estimator. 

\section{Discussion}\label{sec:discuss}

In this paper, we address the problem of how to learn parameter posteriors for ``fully-observed'' \glspl{abm}, that is, when the full trace of the agents' states and interactions are observed. We propose the use of temporal graph neural networks in neural \gls{sbi} algorithms as a way to incorporate useful inductive biases reflecting the natural dynamic graph structure of  \glspl{abm}. Through experiments performed on an \gls{abm} modelling the coevolution of agent opinions and relationship strengths in a dynamic social network, we demonstrated that such an approach can generate approximate Bayesian parameter posteriors in which the ground-truth parameter is assigned a high-posterior density, suggesting that the approximate posterior is accurate to some degree. In future work, we will conduct a more thorough assessment of the quality of the estimated posteriors following the guidelines discussed in \citet{dyer2022black}, for example through the use of posterior predictive checks to assess the predictive power of the inferences drawn, or simulation-based calibration \citep{talts2020validating} to assess the quality of the overall shape of the posterior. In addition, we will extend this approach to continuous-time settings through the use of architectures that are compatible with continuous-time data \citep[see e.g.][]{rossi2020temporal}.

\section*{Acknowledgements}
The authors thank the anonymous reviewers for their helpful comments and feedback. JD is supported by the EPSRC Centre For Doctoral Training in Industrially Focused Mathematical Modelling (EP/L015803/1) in collaboration with Improbable.



\nocite{langley00}

\bibliography{references}
\bibliographystyle{icml2022}

\newpage
\appendix
\onecolumn
\section{Further experimental details}

The first module in the embedding network $g_{\phi}$ is a graph convolutional gated recurrent unit proposed in \citep{seo2018structured}, in which we use $Q=3$ Chebyshev coefficients in the graph filtering operation and choose a hidden state size of 64, such that the hidden state of each agent is a 64-dimensional vector. A single linear layer reduces this $N \times 64$ matrix into an $N$-vector, where $N$ is the number of agents in the system. An embedding of the entire graph then proceeds by passing this $N$-vector through a feedforward network with layer sizes $32, 16, 16$. In our experiments, we take $N = 20$ and simulate for $T = 25$ time steps.

To construct the posterior estimator, we use a masked autoregressive flow \citep{papamakarios2017masked} with 5 transforms and 50 hidden features.

To train the neural networks described above, we follow e.g. \citet{Lueckmann2017, Greenberg2019}; and \citet{dyer2021deep} and train the parameters for the embedding network and the posterior estimator concurrently on the same loss function (the sample mean of the the negative log-likelihood of the parameters). We use a learning rate of $5 \times 10^{-4}$ and a training batch size of 50. Furthermore, we reserve $10\%$ of the training data for validation and cease training if the validation loss fails to decrease after 20 epochs to prevent overfitting. Throughout, we make use of the \texttt{sbi} \citep{tejero-cantero2020sbi} and PyTorch Geometric Temporal \citep{rozemberczki2021pytorch} \texttt{python} packages.


\end{document}